# Unusual ferromagnetic band evolution and high Curie temperature in monolayer 1*T*-CrTe$_2$ on bilayer graphene

Kyoungree Park[1†], Ji-Eun Lee[2†], Dongwook Kim[3], Yong Zhong[4,5], Camron Farhang[6], Hyobeom Lee[1], Hayoon Im[7], Woojin Choi[1], Seha Lee[1], Seungrok Mun[1,8], Kyoo Kim[9], Jun Woo Choi[10], Hyejin Ryu[10], Jing Xia[6], Heung-Sik Kim[1], Choongyu Hwang[7], Ji Hoon Shim[3], Zhi-Xun Shen[4,5*], Sung-Kwan Mo[2*], and Jinwoong Hwang[1*]

[1]Department of Semiconductor Physics and Institute of Quantum Convergence Technology, Kangwon National University, Chuncheon, Korea.

[2]Advanced Light Source, Lawrence Berkeley National Laboratory, Berkeley, CA, USA.

[3]Department of Chemistry, Pohang University of Science and Technology, Pohang, Korea.

[4]Geballe Laboratory for Advanced Materials, Department of Physics and Applied Physics, Stanford University, Stanford, CA, USA.

[5]Stanford Institute for Materials and Energy Sciences, SLAC National Accelerator Laboratory, Menlo Park, CA, USA.

[6]Department of Physics and Astronomy, UC Irvine, Irvine, CA, USA.

[7]Department of Physics, Pusan National University, Busan, Korea.

[8]Interdisciplinary Program in Earth Environmental System Science and Engineering, Kangwon National University, Chuncheon, Korea.

[9]Korea Atomic Energy Research Institute, Daejeon, Korea

[10]Center for Semiconductor Technology, Korea Institute of Science and Technology, Seoul, Korea.

† equal contribution

* Corresponding authors: zxshen@stanford.edu, skmo@lbl.gov, jwhwang@kangwon.ac.kr

2D van der Waals ferromagnets hold immense promise for spintronic applications due to their controllability and versatility. Despite their significance, the realization and in-depth characterization of ferromagnetic materials in atomically thin single layers, close to the true 2D limit, has been scarce. Here, a successful synthesis of monolayer (ML) 1*T*-CrTe$_2$ is reported on a bilayer graphene (BLG) substrate via molecular beam epitaxy. Using angle-resolved photoemission spectroscopy and magneto-optical Kerr effect measurements, that the ferromagnetic transition is observed at the Curie temperature ($T_C$) of 150 K in ML 1*T*-CrTe$_2$ on BLG, accompanied by unconventional temperature-dependent band evolutions. The spectroscopic analysis and first-principle calculations reveal that the ferromagnetism may arise from Goodenough-Kanamori super-exchange and double-exchange interactions, enhanced by the lattice distortion and the electron doping from the BLG substrate. These findings provide pivotal insight into the fundamental understanding of mechanisms governing 2D ferromagnetism and offer a pathway for engineering higher $T_C$ in 2D materials for future spintronic devices.

Layered van der Waals (vdW) ferromagnets have attracted great attention because they provide a versatile playground to explore the nature of low-dimensional magnetism and a rich material basis for future applications.[1-6] The possibilities of controlling the magnetic interaction through external stimuli and constructing various heterostructures open up viable routes for spintronic applications and exploration of novel quantum phenomena.[3-6] However, a critical challenge hindering the practical application of 2D vdW ferromagnets is their low Curie temperature ($T_C$).[3,6] Extensive efforts have been focused on enhancing $T_C$ above room temperature, employing strategies such as chemical doping, strain engineering, and heterostructure design.[4-6]

To engineer $T_C$ of 2D ferromagnets, in principle, a thorough examination of the electronic structure is crucial since it provides direct information on the nature of magnetic interactions in ferromagnetic materials.[7-23] The changes in the electronic structure during a ferromagnetic transition reflect the nature of the magnetism, whether it is mediated by itinerant electrons or direct interaction of local spins.[7-9] It further provides information on the driving factors of ferromagnetism by revealing doping levels, charge carrier density, and correlation effects.[11-19] Precise measurements of the electron band structure and its temperature evolution can unveil the underlying mechanisms of ferromagnetism and help develop designer 2D ferromagnetic systems with higher $T_C$.[14,17,20-30]

In this context, layered 1$T$-CrTe$_2$ stands out as an exceptional material platform that provides insight into higher ferromagnetic transition temperatures. 1$T$-CrTe$_2$ is a member of the transition metal dichalcogenide (TMDC) family, providing advantages in developing the vdW heterostructure of engineered properties. It has an above-room-temperature $T_C$ in both its bulk and thin film forms,[30-32] and displays excellent controllability of its magnetic properties through external stimuli such as charge doping and strain.[32-40] For example, tensile strain on ML 1$T$-CrTe$_2$ can alter its ground states from a ferromagnet to an antiferromagnet.[32-34] In addition, the transition temperature depends significantly on the substrates: The $T_C$ of ML 1$T$-CrTe$_2$ on 1$T$-ZrTe$_2$ and bilayer graphene (BLG) have been reported as 150 – 200 K, among the highest for the single layers of 2D vdW ferromagnets,[32,35] whereas the $T_C$ drops to 50 K in ML 1$T$-CrTe$_2$ on Si(111).[34] Nonetheless, the $T_C$ of ML 1$T$-CrTe$_2$ remains relatively high defying the widely observed dramatic drop in $T_C$ in other 2D vdW magnets, e.g., Fe$_3$GeTe$_2$ ($T_{C,\,bulk}$ = 180 K → $T_{C,\,ML}$ = 20 K).[36]

Despite these advantages, the inherent tendency of 1$T$-CrTe$_2$ to prefer forming Cr-intercalated structures (Cr$_{1+x}$Te$_2$) or other structural phases, such as Cr$_2$Te$_3$ or Cr$_2$Te$_5$, poses challenges in achieving pure 1$T$-CrTe$_2$ crystal[30] and inhibits comprehensive investigations of electronic structure. Such experimental challenges have created a significant barrier against a comprehensive understanding of the microscopic mechanisms underpinning the high-temperature ferromagnetism in ML 1$T$-CrTe$_2$.

In this work, we report a successful molecular beam epitaxy (MBE) growth of ML 1$T$-CrTe$_2$ on a bilayer graphene (BLG) substrate and an investigation of its electronic and magnetic properties across the ferromagnetic transition. Using in situ angle-resolved photoemission spectroscopy (ARPES) and magneto-optical Kerr effect (MOKE) measurements, we observe a ferromagnetic transition with $T_C \approx$ 150 K, which is accompanied by temperature-dependent band shifts and incoherent-to-coherent band crossover. Core-level measurements and density functional theory (DFT) calculations further reveal that the ferromagnetism is closely linked to the changes in the lattice degree of freedom and charge transfer from the BLG substrate, which suggest Goodenough-Kanamori (GK) super-exchange and double-exchange as potential mechanisms behind the ferromagnetism. Our findings establish the epitaxially grown ML 1$T$-CrTe$_2$ on BLG as a premier materials platform to understand the characteristics of vdW ferromagnets in 2D limit.

**Results**

Monolayer (ML) 1$T$-CrTe$_2$ film was synthesized by molecular beam epitaxy (MBE) on a bilayer graphene (BLG) substrate, as schematically illustrated in Figure 1**a,b**. Figures 1**c,d** present distinct reflection high-energy electron diffraction (RHEED) images of MBE-grown sub-ML Cr$_{1+x}$Te (high coverage) and CrTe$_2$ (low coverage), respectively. Since 1$T$-CrTe$_2$ is not an energetically favored phase in Cr-Te compounds,[14,30,41] we find that ML 1$T$-CrTe$_2$ can be synthesized, without other phases, only when the coverage is less than 0.5 ML in a high Te flux environment (Note S1 and Figure S1b, Supporting Information). We further notice that higher coverage films tend to form Cr$_2$Te$_3$ rather than 1$T$-CrTe$_2$ (Note S1 and Figure S1b, Supporting Information). Sharp vertical RHEED line profiles in Figure 1**d** demonstrate the formation of ML 1$T$-CrTe$_2$ film with a well-defined structure, with the line spacing separating it from other structural phases. Using the lattice constant of BLG as a reference, we can estimate the lattice constant of ML 1$T$-CrTe$_2$ on BLG to be ≈ 3.81 Å, which is comparable to the bulk value ( ≈ 3.79 Å) and consistent with previous works.[30-32] The angle-integrated core level photoemission spectrum (Figure 1**e**) exhibits the characteristic peaks of Cr 3$p$ and Te 4$d$, indicating the pure 1$T$-CrTe$_2$ films without a mixture of any other structural phases, such as CrTe and Cr$_2$Te$_3$.[14,34,42]

Figures 1**f** presents the in situ ARPES intensity map ML 1$T$-CrTe$_2$ on BLG substrate taken along the Γ – K direction using $s$-polarized 55 eV photons at 15 K. Characteristic band structures of 1$T$ TMDC are observed at Γ point, where two hole bands of Te 5$p_{x,y}$ character are near Fermi energy ($E_F$) and an M-shape of Te $p_z$-derived band is located at 1.2 eV below $E_F$.[43-48] At K point, the bottom of an electron band from Cr $e_g$ orbital is right at $E_F$.[38] The overall ARPES band structure is in agreement with the DFT+$U$ calculations (Figure 1g and Note S2, Supporting Information). The Fermi surface of ML 1$T$-CrTe$_2$ is made of small hole and electron pockets at the Γ and the K points, respectively. It also demonstrates the absence of multiple domains from azimuthal disorder in our high-quality film (Figure 1**h**).[49-53] The magnetic property of ML 1*T*-CrTe$_2$ film is characterized by magneto-optical Kerr effect (MOKE) measurement as shown in Figure 1**i**. With decreasing temperature, finite Kerr rotations appear 150 K depending on the directions of the magnetic field, indicating that a ferromagnetic phase transition occurs at Curie temperature ($T_C$) ≈ 150 K in ML 1*T*-CrTe$_2$ film grown on BLG substrate.

    A thorough investigation on the temperature evolution of the low-energy electronic structure was made as shown in Figure 2**a**. The ARPES data were taken along the M–Γ–M direction with *s*-polarized 55 eV photons. The obtained ARPES intensity maps clearly show that both inner and outer Te 5$p_{x,y}$ bands have a significant shift with increasing temperatures. To track the movements of Te 5$p_{x,y}$ bands, we plot temperature-dependent (i) positions of leading edges of the inner band taken from energy distribution curves (EDCs) and (ii) the distance between the momentum distribution curve (MDC) peaks at $E_F$ ($\Delta k_x$) in Figure 2**b,c**, respectively. The inner band continuously shifts toward $E_F$ by 200 meV from 10 K to 300 K (Figure 2**b**). Simultaneously, the outer hole band also moves upwards (Figure 2**c**), as evidenced by the gradually increasing $\Delta k_x$ with elevating temperatures. We further notice that the shift rate of these bands is much higher 150 K as denoted by black dashed ovals in Figure 2**b,c**, which coincides with the $T_C$ determined by MOKE measurement (Figure 1**i**).

    For deeper insights on the ferromagnetic transition in ML 1*T*-CrTe$_2$ on BLG, we performed in situ polarization-dependent ARPES measurements up to 4 eV binding energy to observe the temperature evolution of Cr 3$d$ $t_{2g}$ bands. Figures 3**a-f,g-l** display the temperature-dependent ARPES band structures measured along the M–Γ–M direction using *s*-polarized and *p*-polarized 55 eV photons, respectively. Two stand-out features with increasing temperatures are: (i) the coherent Cr 3$d$ bands at Γ and M points, ranged from 1.0

to 2.0 eV below $E_F$, become incoherent with increasing temperature (Figure 3**a-f**), and (ii) the clear splitting between Te $5p_z$ and Cr $3d$ bands, as large as 380 meV and located at 1.3 eV below $E_F$ (white arrows in Figure 3**g-l**), merges together at elevated temperature. Temperature cycling has been performed to eliminate the possibility of sample surface contamination during the heating and cooling processes (Note S3 and Figure S6, Supporting Information).[54]

Figure 3**m** summarizes the temperature-dependent band evolutions of the Te $5p_z$ – Cr $3d$ splitting size around the Γ point (purple), and the peak position (red) and its full width at half maximums (FWHMs) (blue) of the Cr $3d$ band extracted from EDCs at the M point. The size of the band splitting, extracted at $k_x = 0.1$ Å$^{-1}$, gradually decreases with increasing temperature from its maximum value of 380 meV. The merge of two bands accelerates 150 K, and it becomes impossible to define two bands clearly for temperatures above 200 K (Figure 3**m**; purple). The position of the Cr $3d$ band at the M point varies more than 220 meV with increasing temperature and also exhibits a change of the shift trend around 150 K (Figure 3**m**; red). We can exclude the possibility of a simple chemical potential change since (i) we observed different magnitudes of shifts for Te $5p_{xy}$ and Cr $3d$ bands as shown in Figures 2**b** and 3**m**, respectively, and (ii) the $e_g$ band at $E_F$ and the bonding state located at 2.5 eV to 3.5 eV below $E_F$ are robust against the increasing temperatures (Note S3 and Figure S7, Supporting Information). The FWHMs of the Cr $3d$ band taken from EDCs at the M point concomitantly increase with elevating temperatures by 640 meV, which is much larger than the expected thermal broadening from electron-phonon scattering in the order of $k_BT$ (25 meV) (Figure 3**m**; blue),[54-57] where $k_B$ is the Boltzmann constant.

**Discussion**

The observed temperature-dependent band evolutions in ML 1$T$-CrTe$_2$ on BLG, summarized in Figures 2 and 3**m**, are induced by the ferromagnetic transition and the $T_C$ aligns well with the one from MOKE measurements (Figure 1**i**). In addition to the shifts of the electronic bands, the spectral linewidth of the quasiparticle peak exhibits a significant increment with increasing temperature (blue color in Figure 3**m**), with a surge of the broadening rate at and above the $T_C$, suggesting the loss of magnetic order that increases electron scattering and reduces the lifetime of the quasiparticles.[17-21]

Although substantial temperature-dependent band shifts associated with long-range magnetic order have been reported in many magnetic systems,[13,14,17-21] the band evolutions observed in ML 1$T$-CrTe$_2$ on BLG with increasing temperatures below and above $T_C$ are quite unusual when compared to the conventional ferromagnets. In ML 1$T$-CrTe$_2$, the electronic band features all show gradual changes with increasing temperatures below and above the transition temperature. It is actually the rate of change that defines the transition temperature, where the band evolution accelerates only around the $T_C$. This is in contrast to the previous studies on ferromagnetic vdW materials. For example, Fe$_3$GeTe$_2$ just exhibits coherent-to-incoherent band crossover with a negligible band shift, which is often associated with the localized Heisenberg picture.[8,17] CrGeTe$_3$ shows a two-step electronic response due to orbital-selective magnetic ordering: Te 5$p$-dominated bands undergo changes at $T_C$ (65 K), while the changes in Cr 3$d$-dominated bands occur at a higher temperature scale (150 K), due to the development of spin correlations through the dimensional crossover of the magnetic order.[18] In contrast, our results show the continuous band shifts of the Te 5$p$ and coherent-to-incoherent transition of the Cr 3$d$ band with increasing temperature (Figures 2 and 3), indicating complex interdependence between the orbital-dependent band evolutions and the ferromagnetic order.

Now, we turn our attention to the underlying mechanisms of ferromagnetism that can be deduced [7-23] from the observed temperature-dependent band evolutions. Given the metallic character of ML 1$T$-CrTe$_2$, a Stoner model, which describes itinerant ferromagnetism, is a good starting point for a qualitative understanding.[58] Indeed, we found the temperature-dependent giant band shift and splitting (Figures 2 and 3) often associated with the Stoner model,[7,10,13,,14,20] and several studies suggest the itinerant ferromagnetic transition in ML 1$T$-CrTe$_2$.[37,59,60] However, from the Stoner criterion $I·D(E_F) > 1$,[58] where $D(E_F)$ is the density of states (DOS) at $E_F$ and $I$ is the exchange parameter, the Stoner model can be ruled out in our system due to a small DOS at $E_F$ (dot-like Fermi surface in Figure 1**h**). At the same time, the observed band shifts (Figures 2 and 3) do not follow the closing of the split bands above $T_C$ that is expected from the Stoner model (Figure S8a, Supporting Information).

As an alternative mechanism, the Goodenough-Kanamori (GK) super-exchange interaction can be considered due to the crystal structure of 1$T$-CrTe$_2$.[11,21,37,61] GK super-exchange model describes the indirect exchange interaction between two nearest-neighboring Cr ions mediated by Te ion. Types of long-range magnetic order, driven by GK super-exchange, depend on a Cr-Te-Cr bond angle: a bond angle of 90° leads to ferromagnetic coupling, whereas an angle of 180° results in antiferromagnetic coupling.[62] Considering the octahedral structure of 1$T$-CrTe$_2$, this may indicate that the bond angle deviates from 90° at higher temperatures and gradually approaches 90° as the temperature reaches $T_C$, as described in Figure 4**a**.[30,61,62] Indeed, the core-level measurements show changes in both Cr 3$p$ and Te 4$d$ peaks with increasing temperatures (Figure 4**b**) with similar trends observed for Te 5$p$ and Cr 3$d$ bands (Fitures 2 and 3), i.e., the energy shifts accelerate at $T_C$ as denoted by black arrows. This indicates the chemical shifts due to the changes in the interatomic coordinates occur continuously in this system.

To further investigate the potential magneto-structural coupling, we have performed a dynamical mean-field theory (DMFT) calculation. Since DFT+$U$ often yields inadequate energetic corrections and oxidation states due to electron occupancy issues, especially in systems with highly covalent Te atoms in correlated metallic systems, DFT+DMFT offers better control over electron occupancy through the double-counting correction, allowing for accurate reconstruction of Cr-related bands. Our DFT+DMFT on ML 1$T$-CrTe$_2$ further supports that the loss of magnetic order is closely related to the lattice distortion in the octahedra structure due to the Cr $t_{2g}$ degeneracy (Note S5, Supporting Information). Concomitantly, the Raman measurement reveals structural modifications that occur concurrently with magnetic ordering (Note S6, Supporting Information), which is aligned to the core-level shift (Figure 4b), supporting the magneto-structural coupling. These results suggest that the lattice distortion in ML 1$T$-CrTe$_2$ establishes an energetically favorable structure in the ferromagnetic phase with the Cr-Te-Cr bond angle of 90°.[33–40,63,64] This observation is consistent with a previous report on bulk CrTe$_2$, in which the Cr-Te-Cr angle approaches 90° as the ferromagnetic order sets in.[30] These findings firmly suggest a strong correlation between the lattice and ferromagnetism in ML 1$T$-CrTe$_2$. The persistency of the band and core-level shifts above $T_C$ may be related to the strong magnetic fluctuation as in many ferromagnetic systems.[17–21]

Despite the great connection between the band evolutions and lattice distortion favoring ferromagnetism through GK super-exchange, it cannot completely explain the high $T_C$ of 150 K in ML 1$T$-CrTe$_2$ on BLG due to the weak exchange strength.[62] Hence, the double-exchange mechanism should also be considered.[6,22,65-67] The double-exchange mechanism describes the ferromagnetic long-range order through virtual electron hopping, governed by the Hund's exchange coupling ($J_H$), which occurs between adjacent magnetic ions with mixed valences mediated by non-magnetic states.[22,65,66] As illustrated in Figure 4c, the itinerant $e_g$ electrons can strengthen the exchange interactions through virtual hopping, leading to an enhancement of $T_C$.[22,65,66] However, to activate the double exchange interaction in ML 1$T$-CrTe$_2$, additional electron is necessary since the pristine ML 1$T$-CrTe$_2$ does not occupy the itinerant $e_g$ states (Note S2, Supporting Information). We found that the BLG substrate transfers a number of electrons to ML 1$T$-CrTe$_2$ as evidenced by direct observation of the hole-doped BLG $\pi$ bands (Note S4 and Figure S10, Supporting Information), which is due to the huge work function difference between BLG and 1$T$-CrTe$_2$ layers (Figures S4 and S9, Supporting Information).[68-70] This brings the Fermi level of ML 1$T$-CrTe$_2$ on BLG substrate upward compared with that expected from the DFT calculation (Figure 1f,g, and Note S2, Supporting Information). As a result, Cr $e_g$ states are partially occupied in ML 1$T$-CrTe$_2$ (electron pocket at the K point in Figure 1f,h), which allows the interplay between itinerant $e_g$ and localized $t_{2g}$ via Hund exchange coupling.[6,22,65,66] Strong contrast between our results and the ML 1$T$-CrTe$_2$ film grown on Si(111) substrate,[34] whose $T_C$ only reaches 50 K, further support this picture. In 1$T$-CrTe$_2$/Si(111), only a negligible electron transfer from Si(111) occurs due to the less significant work function difference between Si and 1$T$-CrTe$_2$.[68,71] Moreover, the electron transfer from the BLG to the 1$T$-CrTe$_2$ layer makes an enhancement of Hund's coupling by filling the Te ligand holes (Note S4, Supporting Information). The deficiency of electrons in the Te $p$-orbitals should suppress the energy gain of intra-site ferromagnetic alignment of spins via Hund's coupling at Te sites, which leads to the weakening of ferromagnetic super-exchange process, since the ferromagnetic super-exchange becomes maximally effective when there is no hole in the anion (Te) $p$-orbitals (Figure S11b,c, Supporting Information). Therefore, the decrease of holes in Te ligands can lead to the enhancement of the Hund driven ferromagnetic covalent exchange. This scenario is also consistent with previous reports on heavily electron-doped bulk CrGeTe$_3$, leading to an enhancement in $T_C$ from 67 K (undoped) over 200 K (doped).[21,22,65,66]

The substrate effects on ML 1$T$-CrTe$_2$ provide a hint to further enhance $T_C$ of vdW magnets in ML limit. Higher $T_C$ in 2D magnets has been intensively sought after through strain, organic intercalation, ionic gating, and alkali metal dosing on the surface,[35,37,65-67,72,73] mostly in bulk and few-layer vdW ferromagnetic materials. Our study provides direct insights for understanding and controlling the properties of 2D vdW ferromagnets, particularly at the ML level: a charge doping and strain[24,37,67] from a smart choice of substrate could present a novel approach to effectively enhance $T_C$.[39,73-75] In addition, our findings highlight the close-knit relationship between the lattice distortion and ferromagnetic transition in the 2D vdW ferromagnetic materials, which translates into unusual temperature-dependent electronic band evolutions. These results not only contribute to a better understanding of vdW ferromagnet, but also offer important clues for the development of next-generation magnetic materials with higher $T_C$.[4-6]

In summary, we have successfully synthesized a ferromagnetic ML 1$T$-CrTe$_2$ on a BLG substrate by MBE. We observe unusual temperature-dependent band evolutions through ferromagnetic transition ($T_C$ = 150 K) in ML 1$T$-CrTe$_2$ on BLG by combined investigation with ARPES, MOKE, and core-level measurements. Our spectroscopic analysis and first-principle calculations show that GK super-exchange and double-exchange interactions, amplified by lattice change and electron doping from the BLG, should be considered to understand the ferromagnetic transition. Our findings provide crucial information for the fundamental understanding of the nature of 2D ferromagnetism and a pathway to enhance $T_C$ of 2D vdW magnets for future spintronic applications.

**Methods**

**Thin film growth and in-situ ARPES measurement**

The ML 1$T$-CrTe$_2$ films were grown by molecular beam epitaxy (MBE) on epitaxial bilayer graphene (BLG) on 6$H$-SiC(0001) and transferred directly into the ARPES analysis chamber for the measurement at the HERS endstation of Beamline 10.0.1, Advanced Light Source, Lawrence Berkeley National Laboratory. The base pressure of the MBE chamber was 3 × 10$^{-10}$ Torr. High-purity Cr (99.99%) and Te (99.999%) were evaporated from standard Knudsen effusion cells. The flux ratio was Cr:Te = 1:30, and the substrate temperature was held at 340 °C during the growth. This yields the growth rate of 80 mins per monolayer monitored by in situ RHEED. After growth, 1$T$-CrTe$_2$ film was annealed at 340 °C for 20 mins with Te flux and for 30 mins without Te flux to improve the film quality. We notice that the coverage of CrTe$_2$ films on BLG is a key factor to grow 1$T$-CrTe$_2$: less than 0.5 ML coverage prefers forming 1$T$-CrTe$_2$ and more than 0.5 ML is Cr$_{1+x}$Te$_2$ (Note S1, Supporting Information). The coverage was determined by the ratio between BLG and CrTe$_2$ components in RHEED data. ARPES data were taken using Scienta R4000 analyzers at the base pressure 3 × 10$^{-11}$ Torr. The photon energy was set at 55 eV for $s$- and $p$-polarizations with energy and angular resolution of 15–25 meV and 0.1°, respectively. To achieve high-quality ARPES data of low coverage ML 1$T$-CrTe$_2$ film, we performed in situ ARPES measurement and take the spectra longer than 30 mins per single intensity map. The spot size of the photon beam on the sample was ≈ 100 × 100 μm$^2$.

**MOKE measurement**

MOKE measurements were performed in an optical windowed cryostat (with external optics outside of the sample space) with magnetic field applied by a permanent magnet outside the vacuum shroud. MOKE data were taken during warm-up (2 K per a min) in zero-field after cooling in a field of 200 – 500 G, and removing the field at base temperature at 4 K.

**Density Functional Theory Calculations**

The density functional theory (DFT) calculations were performed using VASP (Vienna Ab initio Simulation Package).[76-80] Structural relaxations were carried out using a 15 × 15 × 1 Γ-centered grid, with a plane-wave energy cutoff set to 400 eV. All calculations were conducted using a revised Perdew-Burke-Ernzerhof generalized gradient approximation of the exchange-correlation potential for solids (PBEsol).[81] Total energy and force criteria of $10^{-8}$ eV and $10^{-4}$ eV Å$^{-1}$ were adopted for the charge self-consistent calculations and lattice optimizations, respectively. For the incorporation of the ferromagnetism and the on-site Coulomb repulsions, a simplified rotationally-invariant flavor of the DFT+$U$ method was employed,[82] where $U_{\text{eff}} = U - J_H = 3$ eV was adopted for the Cr $d$-orbital ($J_H$ being the Hund's coupling term).

**Acknowledgements**

K.P. and J.-E.L. contributed equally to this work. The work at the SIMES/Stanford is supported by the U.S. Department of Energy, Office of Basic Energy Sciences, Division of Materials Sciences and Engineering, under Contract No. DE-AC02-76SF00515. This research used resources of the Advanced Light Source, which is a DOE Office of Science User Facility under contract no. DE-AC02-05CH11231. J.-E. L. was supported in part by an ALS Collaborative Postdoctoral Fellowship. J.H. acknowledges financial support from the National Research Foundation of Korea (NRF) grant funded by the Korean government (MSIT) (RS-2023-00280346), and the GRDC (Global Research Development Center) Cooperative Hub Program through the NRF funded by the Ministry of Science and ICT (MIST) (RS-2023-00258359).

C.H. acknowledges financial support from NRF grant funded by the Ministry of Science and ICT (RS-2023-00221154) and NFEC grant funded by the Ministry of Education (RS-2021-NF000587), the Korea Basic Science Institute (National Research Facilities and Equipment Center) grant funded by the Ministry of Education (RS- 2024-0043534), and PNU-RENovation(2023-2024). H. R. acknowledges financial support from the KIST Institutional Program (2E33581) and the NRF (2020R1A5A1016518) grants by the MSIT. S. M. and H.-S. K thank the support of the Basic Science Research Program through the National Research Foundation of Korea funded by the Ministry of Science and ICT [Grant No. NRF-2020R1C1C1005900, RS-2023-00220471]. K.K. was supported by the internal R&D program at KAERI (grant 524550-25) and NRF (2016R1D1A1B02008461).

**Author contributions**

J.W.H. and S.-K.M. initiated and conceived the research. J.W.H., Y.Z., K.P., H.L., H.I., W.C., S.L., and J.-E.L. performed the thin film growth and ARPES measurements under the supervision of S.-K.M., Z.-X.S., and C.G.H.. J.W.H analyzed the ARPES data. C.F. carried out MOKE measurements and analyses under the supervision of J.X.. J.-E.L. conducted Raman measurements and analyses under the supervision of J.W.C., J.W.H. and H.R.. D.K, S.M., H.-S.K. and J. H. S. performed DFT calculations and theoretical analyses. K.P., J.-E.L., Z.-X.S., J.W.H. and S.-K.M. wrote the manuscript with the help from all authors. All authors contributed to the scientific discussion.

**Competing interests**

The authors declare no competing interests.

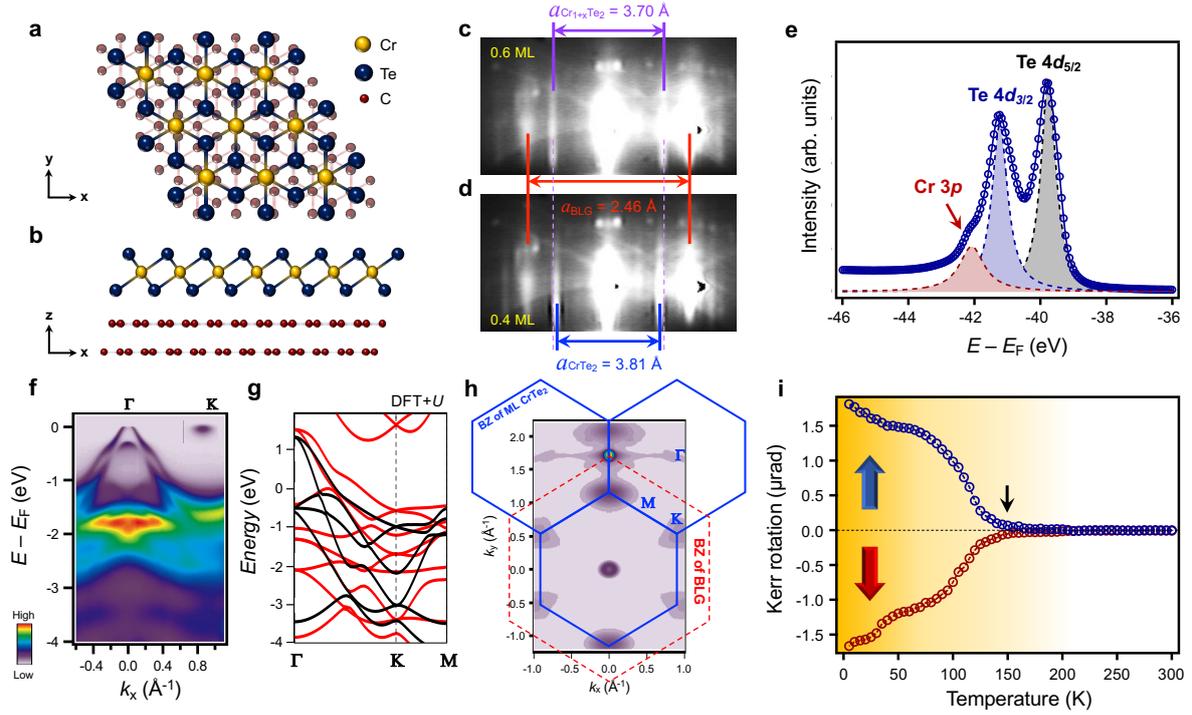

**Figure 1. Characterization of the epitaxial grown ML 1$T$-CrTe$_2$**. (**a,b**) Schematic illustrations of (**a**) top and (**b**) side views of crystal structure of ML 1$T$-CrTe$_2$ on BLG substrate. Yellow, navy blue and red balls represent Cr, Te, and C atoms, respectively. (**c,d**) RHEED images of MBE-grown (**c**) Cr$_{1+x}$Te$_2$ (0.6 ML coverage) and (**d**) CrTe$_2$ (0.4 ML coverage). (**e**) Core level spectra of ML 1$T$-CrTe$_2$ measured at 15 K using 90 eV photons. Red, blue, and black envelopes exhibit deconvoluted peaks of Cr 3$p$, Te 4$d_{3/2}$, and Te 4$d_{5/2}$, respectively. (**f**) ARPES intensity maps of ML 1$T$-CrTe$_2$ on BLG taken along Γ–K direction at 15 K using $s$-polarized 55 eV photon. (**g**) Calculated band structure of freestanding ML 1$T$-CrTe$_2$ by DFT+$U$ method (see Note S2, Supporting Information). Red and black colors indicate the spin majority and minority, respectively. (**h**) Fermi surface ($E - E_F = 0$ eV) of 1$T$-CrTe$_2$ on BLG measured by ARPES at 15 K. Blue and red hexagons indicate the Brillouin zone (BZ) of 1$T$-CrTe$_2$ and BLG, respectively. (**i**) Temperature-dependent Kerr rotation of ML 1$T$-CrTe$_2$ on BLG measured by MOKE. Blue and red arrows present directions of out-of-plane magnetic field. Black arrow indicates the Curie temperature ($T_C$) of ML 1$T$-CrTe$_2$ on BLG.

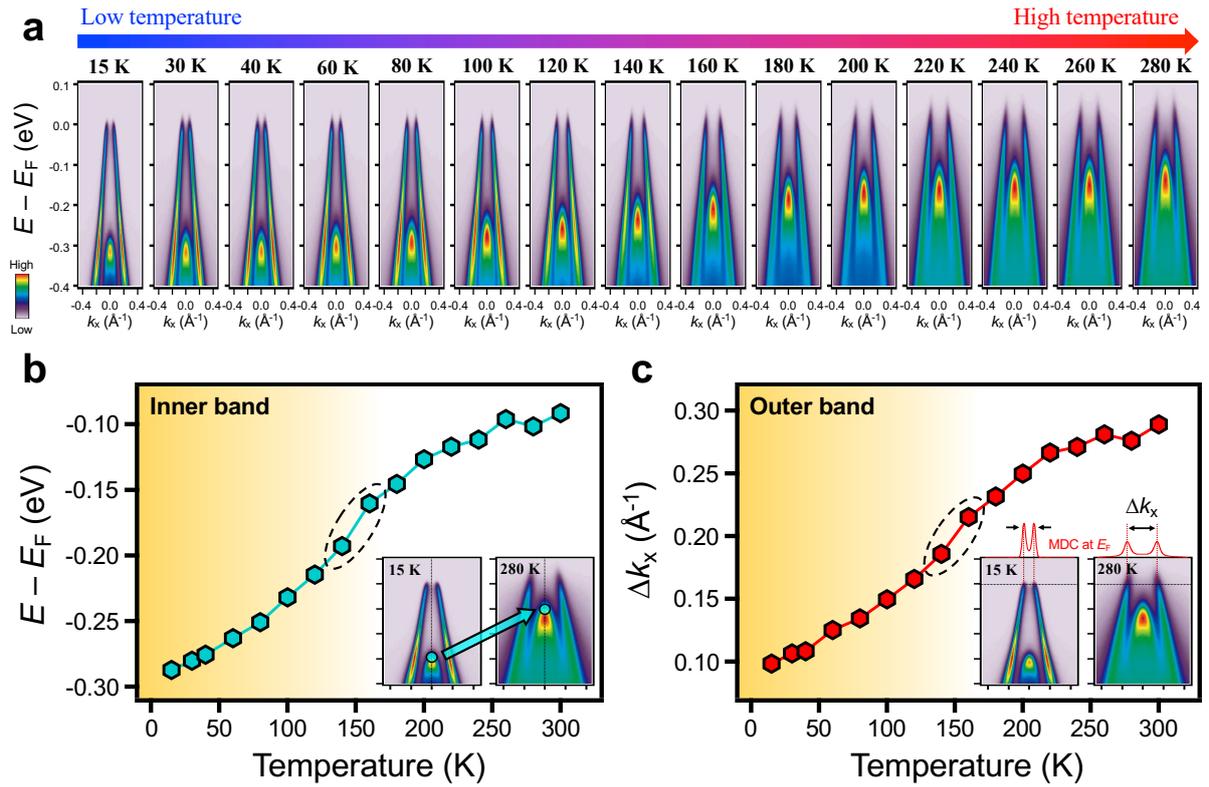

**Figure 2. Temperature-dependent Te $5p_{x,y}$ band evolutions**. (**a**) Band dispersions of Te $5p_{x,y}$ measured along M–Γ–M from 15 K to 280 K using *s*-polarized 55 eV photons. (**b**) Temperature-dependent leading-edge midpoints extracted from EDCs of the inner band as shown in inset. (**c**) Interval of MDC peaks at $E_F$ (as shown in inset) plotted as a function of temperature. The dashed ovals in Figs. 2**c,d** highlight a drastic change of the temperature-dependent behaviors.

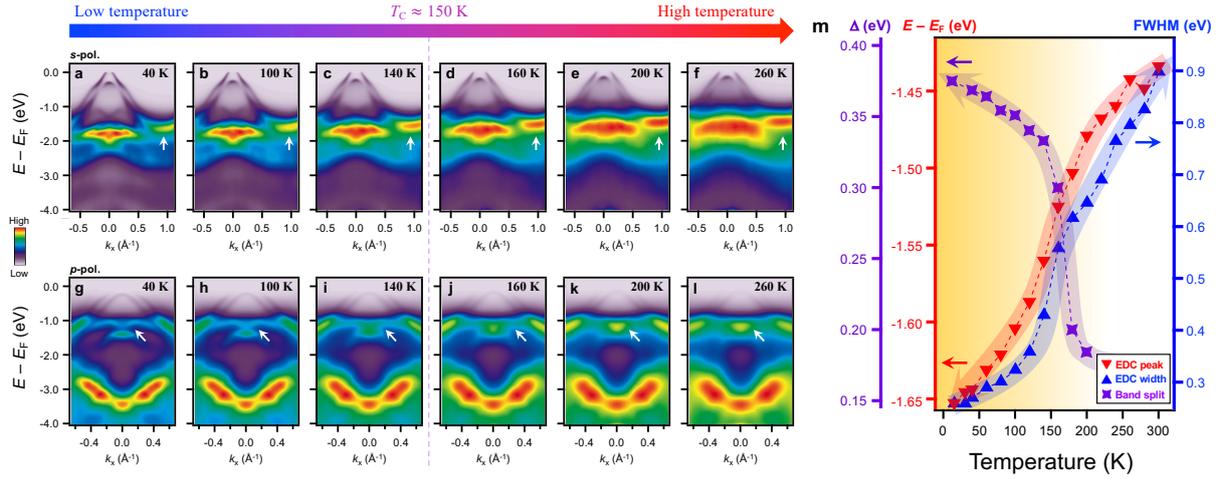

**Figure 3. Polarization- and temperature-dependent ARPES intensity maps of ML 1$T$-CrTe$_2$.** (**a-l**) ARPES intensity maps of ML 1$T$-CrTe$_2$ on BLG taken along Γ−M direction using (**a-f**) $s$- and (**g-l**) $p$-polarized 55 eV photons as a function of temperature. Light purple dashed line presents $T_C$. (**m**) Summary of the evolutions of the size of band split between Te 5$p_z$ and Cr $t_{2g}$ at $k_x$ = 0.1 Å$^{-1}$ (purple), and the peak position (red) and its FWHM (blue) of Cr 3$d$ band extracted from EDCs at M-point, as denoted by white arrows, with increasing temperature.

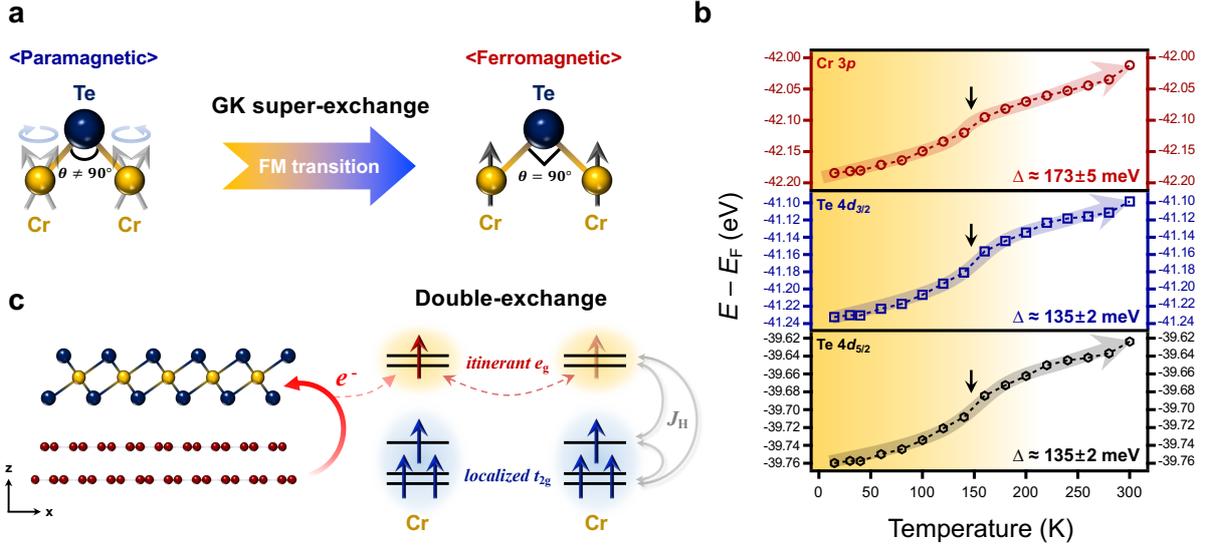

**Figure 4. Possible mechanisms of a novel ferromagnetic transition in ML 1$T$-CrTe$_2$ on BLG**. (**a**) Schematic illustrations of GK super-exchange interaction promoted by lattice rearrangement in ML 1$T$-CrTe$_2$. (**b**) Temperature-dependent core-level peak positions of Cr 3$p$ (red), Te 4$d_{3/2}$ (navy blue), and Te 4$d_{5/2}$ (black). The black arrows in Fig. 4b indicate a point where a huge change of temperature-dependent peak shifts occurs. (**c**) Schematics for double-exchange facilitated by electron transfer from BLG substrate to ML 1$T$-CrTe$_2$ film. $J_H$ represents Hund exchange coupling.

26